# Governance on Social Media Data: Different Focuses between Government and Internet Company

Yu Wenting[1], Shen Fei[2], Min Chen[3]

[1]PhD Candidate, City University of Hong Kong
[2]Associate Professor, City University of Hong Kong
[3]PhD Candidate, City University of Hong Kong, Huazhong University of Science & Technology

A rising number of government requests for personal data have been sent to social media platforms like Facebook in the past few years, which has aroused public concern about information rights and privacy (Musil, 2017; Wired, 2017; Lefebvre, 2018). Although many governments have rules and laws on user data protection and Internet companies continue to claim that their privacy protection program is implementing, there is a school of thought that increasing support of regulations on social media data leads to the legitimization of surveillance (Ingram, 2018).

Both the governments and Internet companies play an important role in governance on social media data. For one thing, governments request for user data to regulate personal speech and behaviors. For another thing, Internet companies choose to respond to government requests or not to regulate their user data. This study uses government requests to Facebook company and Facebook's response rate as an example, to figure out what are the factors related to user data governance of governments and Facebook company. It should be noticed that government requests sent to Facebook include requests for user data from all services provided by the Facebook company, such as Facebook, Instagram and Messenger.

In this study, we intend to answer these questions: What kind of countries send more requests for Facebook user data? What kind of countries get more requests replies from Facebook? Since governments and Internet companies have difference goals in Internet governance, we assume that they would have different motivations on the issue of user data requests.

We listed the potential factors, which include economic, political and social factors. For economic factors, we have GDP per capita in 2017 in US dollars and Internet use percentage. We thought about adopting Facebook market share from StatCounter Global Stat, however, it only provides data of 41 countries in our samples. Hence, we dropped that index. As for political factor, we have political stability and absence of violence. We also consider the possible effects of social factors like personal freedom in a country, and to what extent a country is ruled by law. Since most Internet giants including Facebook are based in the US, the relationship with the US may also influence a country's tendency to send requests to Facebook as well as Facebook's willingness to respond. Hence, we ranked to what extent a country is considered as "friendly" to the US government, to measure the diplomatic relations between the US and other countries. Although the ranking can only represent the public's perceived relationship between a foreign country and the US, it is the best measurement we can find to represent a country's diplomatic relationships with the US. Based on previous research, we listed our research hypotheses as follows:

H1a: Countries with higher GDP per capita request for more user data from Facebook.

H1b: Facebook is more likely to reply to user data requests sent by countries with higher GDP per capita.

H2a: Countries with higher Internet use percentage request for more user data from Facebook.

H2b: Facebook is more likely to reply to user data requests sent by countries with higher Internet use percentage.

H3a: Countries with lower level of political stability request for more user data from Facebook.

H3b: Facebook is more likely to reply to user data requests sent by countries with lower level of political stability.

H4a: Countries with lower level of human freedom requests for more user data from Facebook.

H4b: Facebook is more likely to reply to user data requests sent by countries with higher level of human freedom.

H5a: Countries with lower level of rule of law request for more user data from Facebook.

H5b: Facebook is more likely to reply to user data requests sent by countries with higher level of rule of law.

H6a: Countries more favored by Americans request for more user data from Facebook.

H6b: Facebook is more likely to reply to user data requests sent by countries more favored by Americans.

*Methodology*

We constructed our dataset based on two main data sources. First, we adopted the data of government requests on user data from Facebook's transparency report. The report covers requests to user data not only on Facebook but other services provided by the Facebook company, including Instagram, Messenger, Oculus and WhatsApp from 2013. We adopted the requests data from 2017 since the national indexes we need have their latest updates in 2017.

Both government requests on Facebook user data and Facebook's reply rate act as our dependent variables. But since government requests on user data might be positively correlated with a nation's population, a government's request times divided by the nation's population is used as our dependent variables. Another problem is, the distribution of the requests number is highly skewed toward the 0 end (M=65458, SD=1461.65, Skewness=8.203, Kurtosis=74.759); and therefore, we used the logged number of requests to make its distribution more symmetric (M=1.80, SD=1.13, Skewness= .370, Kurtosis= -.478).

Next, we adpted the indices of GDP per capita, Internet use, political stability and absence of violence, human freedom, rule of law and Americans' favorite countries ranking for the year 2017 from the website of different institutions (see Table 1 for details). We then input the data matching the countries we have.

[Insert Table 1 about here]

Only countries that have sent at least one request for user data to Facebook are included in the sample. Also, countries lack information in more than three indices are deleted from the sample. Finally, the sample includes government requests on user data from 105 countries. We run two OLS regression models to test the effect of listed factors on government requests and Facebook reply rate.

*Results*

The results of OLS regression are listed in Table 2. The value of R-square shows that our listed factors have strong explanatory power. In model 1, requests for Facebook user data act as the dependent variable. According to the results, only H1a and H6a are supported. GDP per capita (B= 1.013, SD = 1.708, p<.001) has a positive while rule of law (B=-3.914, SD= 1.696, p<.05) has a negative impact on requests number. Interestingly, the reverse of H5a is supported, that human freedom (B= .769, SD= .242, p<.01) share a positive relationship with the number of user data requests. Internet use, political stability and absence of violence and to what extent a country is favored by Americans have no significant effect on government requests for Facebook user data.

[Insert Figure 1 about here]

[Insert Table 2 about here]

In model 2, Facebook's reply rate to government requests act as a dependent variable. H3b is supported, that political stability and absence of violence have a negative effect on the dependent variable, which means the Facebook company is more likely to reply to requests from unrest countries. Human freedom shares a positive relation with Facebook's reply rate as we hypothesized in H6b, but Internet use, GDP per capita, rule of law and whether a country is favored by Americans have no significant effect in the model.

*Conclusions*

Our results show that governments and Internet companies like Facebook have different focuses when regulating user data from social media. Countries with a better economy, higher level of political stability and social safety, and lower level of rule of law are more likely to send requests for Facebook user data.

However, a country's GDP per capita and rule of law do not affect the rate of Facebook responses. Facebook company replies to government requests based on human freedom as well as political stability and absence of violence. In addition, Internet use and to what extent a country is favored by US people neither impact government requests nor Facebook's reply rate. The results of this study require further discussions.

Table 1. Descriptive data of indices adopted in this study

| Index | Source | Scale/Unit | M | SD |
|---|---|---|---|---|
| GDP per capita (logged before analysis) | World Development Indicators (The World Bank, 2017) | US dollars | 19144.10 | 21884.88 |
| Internet use | World Development Indicators (The World Bank, 2017) | 0-100% | 63.78 | 24.56 |
| political stability and absence of violence/terrorism | Political Stability and Absence of Violence/Terrorism Index (The World Bank, 2017) | -2.5-2.5 | -.03 | .91 |
| human freedom | Human Freedom Index (Ian & Tanja, 2017) | 0-10 | 7.23 | 1.04 |
| rule of law | Rule of Law Index (The World Justice Project, 2017) | 0-1 | .59 | .15 |
| Americans' favorite countries ranking | American's Friend and Economy (YouGov, 2017) | 1-142 | 59.12 | 39.46 |

Note: We rank the percentage of American's friendly attitude to other countries. The country ranks first is coded as "1".

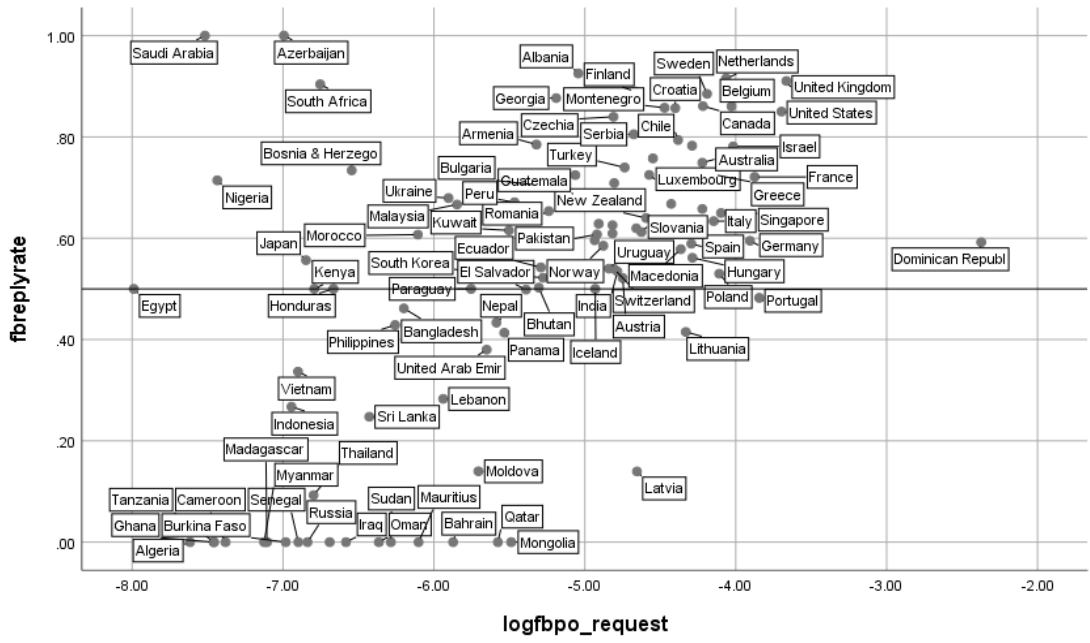

Figure 1.Plot graph of logged requests per capital on Facebook and Facebook's reply rate

Table 2. Predicting user data requests to Facebook, Microsoft and Google

|  | Model 1 | | Model 2 | |
|---|---|---|---|---|
|  | DV: Requests for FB User Data | | DV: FB Reply Rate | |
|  | Unstandardized coefficients | SD | Unstandardized coefficients | SD |
| Intercept | -13.61*** | 1.71 | -1.09* | .43 |
| Log GDP Per Capital | 1.01* | .01 | .12 | .13 |
| Internet Use | .01 | 0.54 | .01 | .00 |
| Political Stability and Absence of Violence | -.22 | .21 | -.15** | .05 |
| Human Freedom | .76** | .24 | .14* | .06 |
| Rule of Law | -3.91* | 1.70 | -.53 | .43 |
| Americans' Favorite Countries Ranking | -0.16 | -0.002 | .06 | .04 |
| $R^2$ | 73.2% | | 64.4% | |

Notes: *p<.05, **p<.01, ***p<.001